\documentclass[twocolumn,showpacs,preprintnumbers,amsmath,amssymb,amsfonts]{revtex4}

\usepackage{graphicx}
\usepackage{dcolumn}
\usepackage{bm}
\usepackage{colordvi}

\begin{document}

\preprint{APS/123-QED}

\title{Shear zones and wall slip in the capillary flow of concentrated colloidal suspensions}

\author{Lucio Isa}
\author{Rut Besseling}%
\author{Wilson C K Poon}%
\affiliation{%
SUPA \& School of Physics, The University of Edinburgh, James Clerk Maxwell Building, The Kings Buildings, Mayfield Road, Edinburgh EH9 3JZ, UK}%

\date{\today}

\begin{abstract}

We image the flow of a nearly random close packed, hard-sphere colloidal suspension (a `paste') in a square
capillary using confocal microscopy. The flow consists of a `plug' in the center while shear occurs localized
adjacent to the channel walls, reminiscent of yield-stress fluid behavior. However, the observed scaling of the
velocity profiles with the flow rate strongly contrasts yield-stress fluid predictions. Instead, the velocity
profiles can be captured by a theory of stress fluctuations originally developed for chute flow of dry granular
media. We verified this behavior both for smooth and rough boundary conditions.

\end{abstract}



\maketitle

Understanding the deformation and flow, or rheology, of complex fluids in terms of their constituents (colloidal
particles, polymer chains or surfactant aggregates) poses deep challenges to fundamental physics, and has wide
industrial applications \cite{LarsonBook}. The experimental study of complex fluid rheology typically starts in a
rheometer, in which stresses and strains are applied and measured in well-defined, `rheometric' geometries
(`cone-plate', etc.). Translating rheometer data to more complex flows is non-trivial, but well developed in
polymers (see, e.g., \cite{McLeishScience}).

The understanding of colloidal flows lags considerably behind and despite their equal practical importance
\cite{LewisCeramics,PasteReview}, studies on model systems have been carried out only recently \cite{haw1}. Compared to polymers, colloids pose some unique challenges. Concentrated
suspensions (`pastes') are generally non-ergodic (or `glassy'), so that {\it any} flow involves non-linearities
(e.g. yielding \cite{petekidis,PhamEPL} or shear thickening \cite{Frith,Lootens}). Moreover, specific geometries
in applications may involve dimensions comparable to single particles and lead to confinement effects, such as in
micro-fluidics \cite{Microfluidics}. The most quantitative theory for quiescent colloidal glasses, mode coupling
theory, has only recently been extended to deal with simple shear \cite{ShearMCT}.

Here we present an experimental study of the flow of a hard-sphere suspension at nearly random close packing, a
`paste', in a twenty-particle-wide square capillary. Pastes are ubiquitous in industry, where their unique
rheology presents many challenges and opportunities \cite{PasteReview}. The simplicity of the geometry is
appealing from the fundamental perspective. It also makes direct contact with microfluidic applications
\cite{Microfluidics}. Using fast confocal microscopy, we tracked the motion of individual colloids and measured
the velocity profiles in channels with both smooth and rough walls. Despite the colloidal nature of our
suspension, we find significant similarities with granular flow, itself of wide applied \cite{GranularReview} and
fundamental \cite{midi} interest.

We used sterically stabilised polymethylmethacrylate (PMMA) spheres of diameter $D = 2.6 \pm 0.1$~$\mu$m (from
confocal microscopy) fluorescently labelled with nitrobenzoxadiazole, and suspended in a mixture of
cyclo-heptylbromide and mixed decalin (viscosity 2.6 mPa$\cdot$s) for buoyancy matching at room temperature
$T=T_r$. Their Brownian time is $\tau_B=D^2/24 D_0 \simeq 4.4$~s with $D_0$ the bare diffusion coefficient.

\begin{figure}[h]
\includegraphics[width=0.4\textwidth,clip]{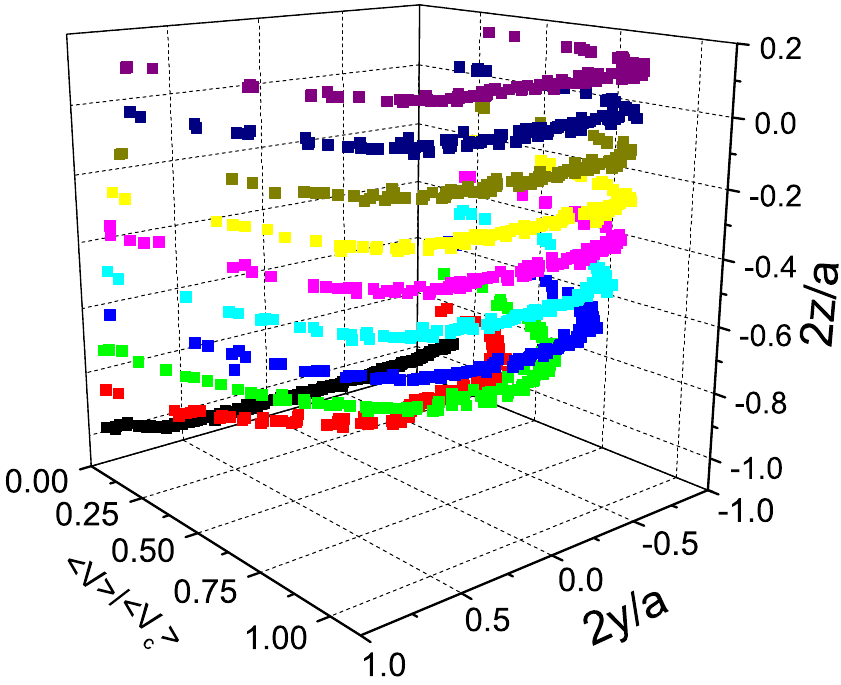}
\caption{Velocity profiles in the lower half of the capillary (width $2a = 50$~$\mu$m) with smooth walls
reconstructed in $3$~$\mu$m steps. The average velocity $\langle V \rangle$ is in units of the average velocity of
the central, unsheared, `plug', $\langle V_{c} \rangle = 20$~$\mu$ms$^{-1}$. \label{fig:vprofile_ff5_251006}}
\end{figure}

A dense suspension (volume fraction $\phi \gtrsim 0.63$, from confocal microscopy) was obtained by centrifuging
(at $T>T_r$ to remove buoyancy matching). A constant pressure gradient, $\nabla p$, is applied to drive the
suspension into a square borosilicate glass micro-channel (Vitrocom Ltd; side $2a = 50$~$\mu$m) \cite{isa1}, whose
inner walls were either untreated and smooth, or coated with a disordered monolayer of colloids and thus rough on
the particle level. The coating particles, which are sightly larger ($\bar{D}_{\rm coat} = 2.8$~$\mu$m) and more
polydisperse PMMA spheres (right inset to Fig. \ref{fig:vfit}), were applied by filling with a dilute suspension
and attached by heating in a vacuum oven (110~$^\circ$C).

\begin{figure}[h]
\includegraphics[width=0.45\textwidth,clip]{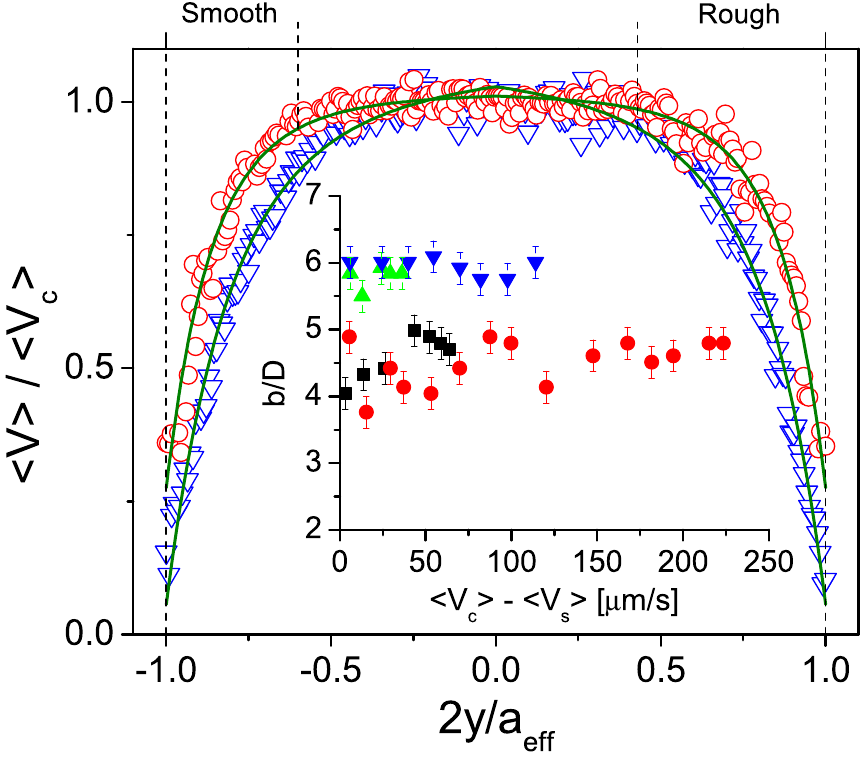}
\caption{Averaged velocity profile as a function of $2y/a_{\rm eff}$ for smooth (\Red{$\circ$}) and rough
(\Blue{$\triangledown$}) walls at $z = 17$~$\mu$m. Full lines: fits from integrating Eq.~\ref{eq:fit}; dotted
lines: extents of sheared zones. $\langle V_{\rm c} \rangle$ is $114$~$\mu$ms$^{-1}$ for the smooth wall case and
$80$~$\mu$ms$^{-1}$ for the rough wall case. Inset: width of shear zone in units of particle diameter as a
function of $\Delta V$. \Green{$\blacktriangle$} and \Blue{$\blacktriangledown$}: two runs for rough walls;
\Red{$\bullet$} and $\blacksquare$: two runs for smooth walls. \label{fig:smoothvsrough}}
\end{figure}

The flow (along $x$) across the full width of the channel ($|y| \leq a$) was imaged with a Visitech VTeye confocal
scanner in a Nikkon TE Eclipse 300 inverted microscope. We collected 44~$\mu$m$\times 58 \mu$m images (107 frames
per second) at depths $-a< z < +0.2a$ ($z=-a$ is the lower surface). From the two-dimensional images, we located
particles with a resolution $\delta x,\delta y \simeq 50$~nm \cite{crocker}. A key step for correct tracking is to
follow colloids in a `comoving' frame \cite{isa1}. This involves removing from the `raw' coordinates the advective
motion $\Delta x(\bar{y})$, which is obtained as the $x$-shift that maximizes the cross correlation between image
strips around $\bar{y}$ in successive frames. From time-dependent coordinates (restored in the lab frame), we
obtained particle velocities \footnote{This procedure applied to the flow of a 30\% suspension in a
two-dimensional channel gave parabolic profiles \cite{isa1} as also found in \cite{frank}.}. After start-up
transients, we found oscillations in the particle velocities (data not shown) at 0.1~Hz (slow flow) to 1~Hz (fast
flow). This feature, which we discuss elsewhere, is ubiquitous in the pipe flow of pastes \cite{yaras,haw1}. Here, we
restrict ourselves to steady-state velocity profiles, $\langle V \rangle$, obtained by averaging over 20~s (slow
flow) to 10~s (fast flow), corresponding to $\sim 2000$ to $\sim 1000$ frames, respectively. We imaged at $x \sim
0.5$~cm from the entrance to the capillary (corresponding to $\sim 2000$ particles), where entry effects have died
out and the results show negligible $x$ dependence.

Typical data for smooth walls are shown in  Fig.~\ref{fig:vprofile_ff5_251006}. For $z \gtrsim 10$~$\mu$m, each
velocity profile consists of a shear zone close to the walls and a nearly unsheared central `plug'. This plug
shrinks at smaller $z$, i.e. closer to the bottom wall. We also observe wall slip with velocity $\langle V_{s}
\rangle$. The profiles for smooth and rough walls, Fig.~\ref{fig:smoothvsrough}, are qualitatively similar, but
the latter displays considerably smaller wall slip and larger shear zones. Note that for rough walls, we use an
effective half width $a_{\rm eff} = a - \bar{D}_{\rm coat} - \bar{D}/2$; for smooth walls, $a_{\rm eff} = a - \bar
{D}/2$.

\begin{figure}[h]
\includegraphics[width=0.45\textwidth,clip]{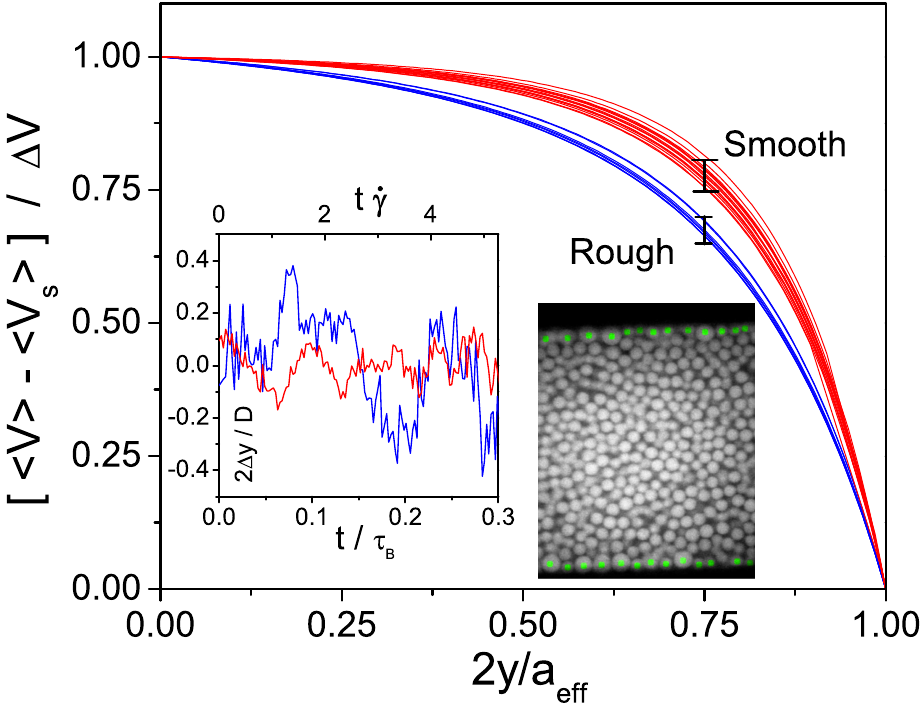}
\caption{Normalized velocity profiles at $z = 17$~$\mu$m versus $2y/a_{\rm eff}$. We use the profiles fitted from
integrating Eq.~\ref{eq:fit}; the error bars show the spread in the measured data. Left inset: normalized
transverse fluctuations of individual particles in the shear zone for the two boundary conditions, versus
normalized time $t/\tau_B$ or {\it local} accumulated strain $t \cdot \partial \langle V(y) \rangle/\partial y=t
\cdot \dot{\gamma}(y)$, top axis. The traces show larger fluctuations in the rough wall case. Right inset: confocal microscopy snapshot of the flow at $z = 17$~$\mu$m for a
channel with rough walls; the particles marked with \Green{$\bullet$} are attached to the walls. \label{fig:vfit}}
\end{figure}

The dependence of the velocity profiles on the overall flow rate is striking. We define the width of the shear
zone $b$ as the distance from the wall where the flow speed is $\langle V_s \rangle + 0.95\Delta V$ where $\Delta
V= \langle V_{\rm c} \rangle - \langle V_{\rm s} \rangle$ is the difference between the averaged center and wall
speeds (with $\Delta V/a$ the average shear rate). We see in the inset to Fig.~\ref{fig:smoothvsrough} that $b$ is
independent of $\Delta V$, i.e. the central plug remains essentially constant for flow rates varying by more than
one decade. Associated with this, the normalized velocity profiles $(\langle V \rangle - \langle V_s
\rangle)/\Delta V$ for different flow rates collapse onto two different master curves for rough and smooth walls,
Fig.~\ref{fig:vfit}.

Let us now compare the observed behavior to what is expected from  bulk rheology. Considering the system as a
glassy yield stress fluid, the stress $\tau$ versus strain rate $\dot{\gamma}$ relation typically displays
Herschel-Bulkley shear thinning behavior: $\tau-\tau_{\rm yield} \propto \dot{\gamma}^{n}$ ($n\approx 0.5$)
\cite{petekidis} with $\tau_{\rm yield}$ the yield stress. Plug flow is ubiquitous in all yield stress fluids, and
has been extensively studied \cite{You05}. Thus, our observation of plug flow is, in itself, unsurprising. In yield stress
fluids though, if the maximum applied shear stress in the channel exceeds $\tau_{\rm yield}$, we have plug flow with
shear zones whose boundaries are given by $\tau = \tau_{\rm yield}$; hence, the width of the shear zone, $b$,
grows with the shear rate and eventually leaves a vanishingly small `plug' \cite{You05}. {\it This is not what we
observe.}

A clue to what may be happening in our experiments comes from the
observation that at the typical strain rate encountered in the shear zones,
$\tau_B(\Delta V/a) \sim 50$, conventional rheology would lead us to expect
severe shear thickening. Shear thickening in pastes is very far from
completely understood on the microscopic level \cite{Frith,cates1,ball}; but we may speculate
that for colloids stabilised by short grafted polymers, interparticle
friction may become important as they jam against each other driven by
shear. Indeed, the measured transverse fluctuations of particles clearly
show the motions induced by them 'bumping' along neighboring layers, Fig. 3
inset. We are thus led to consider analogies with friction-dominated
granular flow.

Indeed, both the width of the shear zone $b \sim 6D$ ($\sim 5D$ for smooth walls) and the lack of flow rate
dependence, are strongly reminiscent of observations in gravity-driven `chute flow' of dry granular materials
\cite{Pouliquen1}. In particular, Pouliquen and Gutfraind observed $b \approx 6 D$ in the two-dimensional flow of
discs (diameter $D$) for channels of widths $10D \leq 2a  \leq 44D$ and developed a model to account for their
observations \cite{Pouliquen1,Pouliquen2}. Below we extend their model to three dimensions, and show that it
indeed predicts $b$ independent of $\Delta V$.

Following \cite{Pouliquen1} we start with the components of the stress tensor {\boldmath$\tau$} in the
fully-developed flow of a continuum medium in the $x$ direction along a pipe, which satisfies
\begin{eqnarray}
\frac{\partial \tau_{xx}}{\partial x} + \frac{\partial \tau_{xy}}{\partial y} + \frac{\partial \tau_{xz}}{\partial z} & = & 0\;, \label{tauxy} \\
\frac{\partial \tau_{yy}}{\partial y} = \frac{\partial \tau_{zz}}{\partial z} & = & 0\;, \label{tauyy}
\end{eqnarray}
with $\partial \tau_{xx}/\partial x = -\nabla p$. For a square pipe of side $2a$ \cite{white}:
\begin{eqnarray}\label{eq:shearstressprofile}
\nonumber
\tau_{xy}(z) =&\tau_{\rm 0}& \displaystyle\sum_{k=1,3,5...}^\infty
(-1)^{(k+1)/2} \times \\
&\times& \left[ 1-\displaystyle{\frac{\cosh (k \pi z/2a)}{\cosh (k \pi
/2)}} \right] \displaystyle{\frac{\sin (k \pi y/2a)}{k^2}}
\end{eqnarray}
where $\tau_{\rm 0} = 8 a \nabla p/ \pi^2$. This constant sets the scale for the ($z$-dependent) maximum stress at
the wall, $\tau_{\rm max}$.

\begin{figure}[h]
\center{\includegraphics[width=0.46\textwidth,clip]{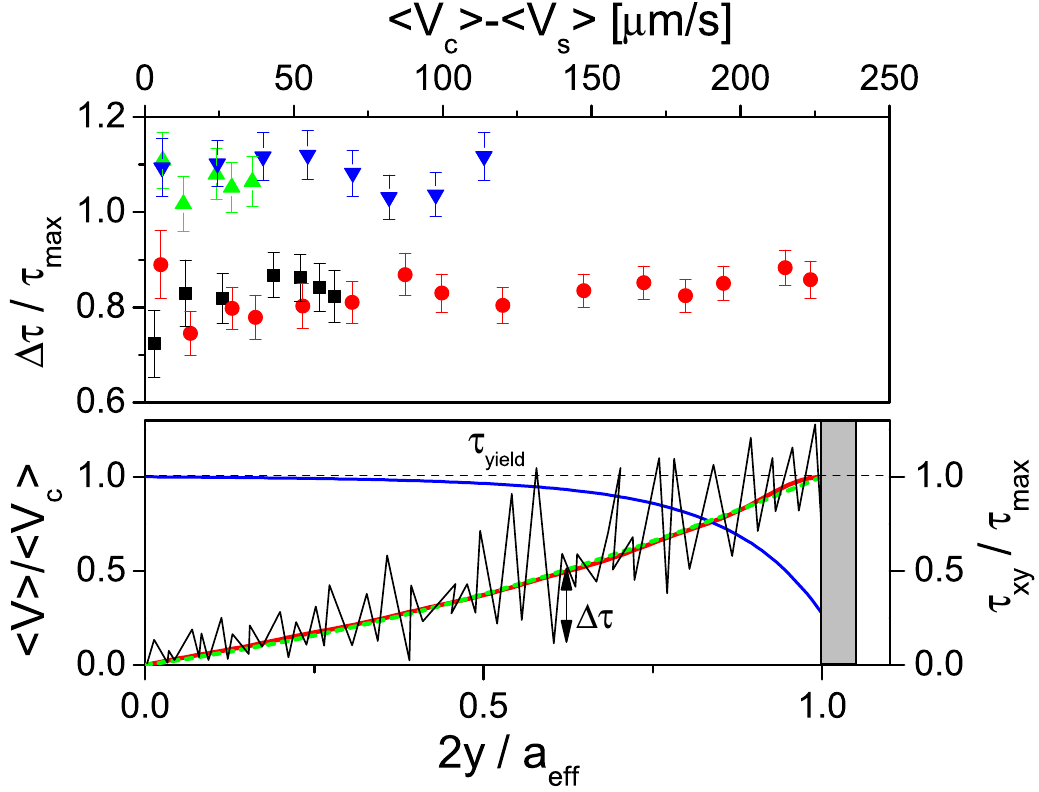}}
\caption{\label{fig:yieldsketch} (Top) Fitted fluctuating stress amplitude $\Delta \tau$ in units of $\tau_{\rm
max}$ as a function of the `net shear', $\Delta V$; symbols as in Fig.~\ref{fig:smoothvsrough}.(Bottom) Schematic
of the fluctuating shear stress superimposed on the continuum $\tau_{xy}$. The dashed line shows the quadratic fit
used for the integration. In the zone where the fluctuating stress overcomes $\tau_{\rm yield}$, flow results.}
\end{figure}

To determine the normal stress, $\tau_{yy}$ we impose a Coulomb friction condition at the walls:
\begin{equation}
\tau_{yy} = \mu_{\rm wall}^{-1} \tau_{\rm max}\;, \label{normal}
\end{equation}
where $\mu_{\rm wall}$ is the friction coefficient between the suspended particles and the wall; this constant
$\tau_{yy}$ satisfies Eq.~\ref{tauyy} \footnote{A completely analogous discussion can be given for $\tau_{xz}$ and
$\tau_{zz}$, leading to a velocity profile $V(z)$.}. Following the Coulomb criterion, the material yields if
$\tau_{xy} \geq \tau_{\rm yield}\equiv \mu_{\rm bulk}\tau_{yy}$, with $\mu_{\rm bulk}$ the friction coefficient
inside the material. We expect $\mu_{\rm wall} \leq \mu_{\rm bulk}$, with `=' for a rough wall coated with
particles, and `$<$' for smooth, glass walls. From Eq.~\ref{normal} we have:
\begin{equation}
\tau_{\rm yield} = (\mu_{\rm bulk}/\mu_{\rm wall})\tau_{\rm max} \geq \tau_{\rm max}\;. \label{tauyield}
\end{equation}
Since $\tau_{xy} \leq \tau_{\rm yield}$, the bulk never yields and the whole material slips as a plug.
Note from Eq.~\ref{tauyield} that $\tau_{\rm yield}$ increases with $\tau_{\rm max}$ (and therefore $\Delta V$). This reflects the rising  normal stress $\tau_{yy}$, Eq.~\ref{normal}, which increases the friction between particles and thus $\tau_{\rm yield}$\footnote{Thus, this {\it variable} Coulombic yield stress is a different quantity from the {\it constant} yield stress featuring in (say) the HB model.}.

We now assume the presence of stress fluctuations which, when added to the continuum $\tau_{xy}$,
Eq.~\ref{eq:shearstressprofile}, may  take the local stress above $\tau_{\rm yield}$, Fig.~\ref{fig:yieldsketch}.
Following \cite{Pouliquen1} we use a `Boltzmann' ansatz for the yielding probability:
\begin{equation}
p_{\rm yield} \propto  \exp \left[- \displaystyle{\frac{ \tau_{\rm yield}(z) - \mid \tau_{xy} (y,z)\mid }{\Delta \tau}} \right]\;, \label{pyield}
\end{equation}
where $\Delta \tau \equiv \Delta \tau(z)$ is the amplitude of stress fluctuations, taken to be independent of $y$
at any particular $z$. The model is completed by the simplest possible ansatz relating shear rate to $p_{\rm
yield}$: $\partial V(y,z)/\partial y \propto p_{\rm yield}$ \footnote{The ansatz \ref{pyield} was proposed
\cite{Pouliquen1} to account for shear-zone behavior next to walls, and is qualitatively incorrect for the center
of the pipe, where we must have $\partial V/\partial y =0$.}. At a fixed $z = z_0$, therefore, we have
\begin{equation}
\frac{\partial V(y,z_0)}{\partial y} = \mbox{constant} \times \exp \left[ \frac{\tau_{xy}(y,z_0)}{\Delta \tau}\right] \label{eq:fit} \;.
\end{equation}
In order to perform the integration analytically we fitted the stress profile, Eq.~\ref{eq:shearstressprofile},
with a quadratic approximation; Fig.~\ref{fig:yieldsketch} (bottom) shows both the analytic expression and the
fit. We then substituted the fitted stress into Eq.~\ref{eq:fit}, integrated and fitted the resulting velocity
profile to the measured data. Example results are shown in Fig.~\ref{fig:smoothvsrough}.

The fitted stress fluctuation amplitudes normalized to the wall stress, $\Delta \tau/\tau_{\rm max}$, at $z=
17$~$\mu$m for both smooth and rough walls are plotted against $\Delta V$ in Fig.~\ref{fig:yieldsketch}. As
$\tau_{\rm max}$ increases with flow rate, $\Delta \tau$ increases proportionally so that their ratio remains
constant. Combined with Eqs.~\ref{eq:shearstressprofile} and \ref{tauyield} this means that
$(\tau_{yield}-|\tau_{xy}|)/\Delta\tau $ in Eq.~\ref{pyield} is independent of $\tau_{\rm max}$, resulting in a
constant value for $b$, Fig.~\ref{fig:smoothvsrough}, and collapse of the velocity profiles, Fig.~\ref{fig:vfit}.

The fact that $\Delta \tau$ scales as the wall stress $\tau_{\rm max}$ suggests that the stress fluctuations are
controlled by what happens at the boundaries. Two dimensional simulations of dry granular matter indeed relate
stress fluctuations to inhomogeneities in the friction-dominated force chains resulting from contacts at the walls
\cite{ShearZoneOrigins}. Here, we find (for $z \sim a$) that $\Delta \tau/\tau_{\rm max}$ for rough walls is
somewhat larger than for smooth walls, Fig.~\ref{fig:yieldsketch}; this difference is directly correlated with the
observed difference in $b$, Fig.~\ref{fig:smoothvsrough}. The presence of such a difference, and its sign, is
unsurprising. Flow along a smooth wall is less `bumpy' and can thus be expected to generate a lower level of
stress fluctuations which propagate through the system, as also shown by the two traces in the left inset to
Fig.~\ref{fig:vfit} \footnote{We also note that for smooth walls we find considerable layering in the structure
close to the walls \cite{isa1}.}.

\begin{figure}[h]
\includegraphics[width=0.45\textwidth,clip]{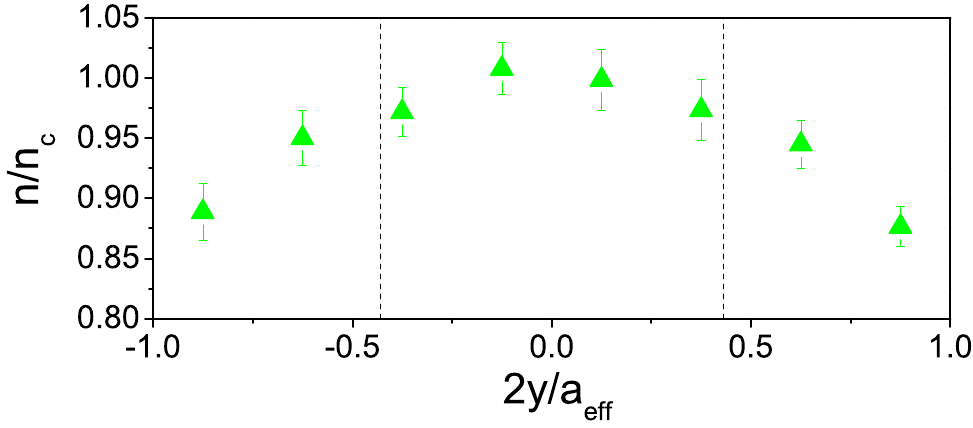}
\caption{Average density profile at $z = 17$~$\mu$m for rough walls, in units of the density $n_c$ in the center.
Dashed lines: average width $b$ of the shear zones in channels with rough walls.} \label{cprofile}
\end{figure}

The analogy with granular flow extends further than the constant shear zone width. In \cite{Pouliquen1}, a
variation of the density across the channel was observed, the `plug' being $\sim 10\%$ denser than the edge of the
shear zone. We observe the same feature in our channel flows. Figure~\ref{cprofile} shows a typical density
profile at $z = 17$~$\mu$m, measured by counting particles per unit area and averaging $10^4$ frames; this profile is essentially independent of the overall flow rate. The observed density reduction in the shear zone is not unexpected,
since dilatancy is required in flowing particulates (dry or wet) at such high packing fractions.

Our findings also have implications concerning the role played by confinement. Since $D/2a \approx
20$, one might expect strong size-dependence. However, the width of the shear zone $b$ we observed, is always smaller than the width of the channel, $a$ and size
effects should only become significant for channels with $\ a \lesssim b$. 

To summarize, we have measured the flow properties of a nearly-random-close-packed hard-sphere colloid driven by a
constant pressure gradient in a twenty-particle-diameter square channel by tracking the motion of individual
particles using fast confocal microscopy. At all flow rates, we observed a central, almost unsheared plug, and
peripheral shear zones. In contrast to the prediction of yield-stress fluid rheology, the size of the shear zone
remained constant as the flow rate was increased. We explained this by appealing to a model originally set up for
the gravity-driven `chute flow' of dry granular materials \cite{Pouliquen1}. This model should be applicable in
our case if the yield stress, $\tau_{\rm yield}$, is dominated by inter-particle friction. The model predicts that stress fluctuations can bring about yielding even
when the average stress is below $\tau_{\rm yield}$. Quantitative fits to our data were obtained.

We thank Andrew Schofield for providing the particles, and Eric Weeks and Alexander Morozov for helpful
discussions. L. Isa was funded by the EU network MRTN--CT--2003--504712, and R. Besseling by EPSRC GR/S10377/01
and EP/D067650/1.

\bibliographystyle{apsrev}
\bibliography{bibliography}

\end{document}